\pdfoutput=1
\documentclass[sigconf]{acmart}

\usepackage{booktabs}
\usepackage{colortbl}
\usepackage{xcolor}
\usepackage{tikz}
\usetikzlibrary{arrows.meta,positioning}
\usepackage{algorithm}
\usepackage{algpseudocode}

\definecolor{keep_green}{HTML}{2ecc71}
\definecolor{q_orange}{HTML}{e67e22}

\makeatletter
\patchcmd{\maketitle}
  {{\itshape \acmConference@shortname, \acmConference@venue}}
  {{\itshape \acmConference@shortname, \acmConference@date, \acmConference@venue}}
  {\typeout{PATCH-OK: permission-block date}}{\typeout{PATCH-FAILED}}
\makeatother

\begin{document}

\acmYear{2026}\copyrightyear{2026}
\setcopyright{cc}
\setcctype[4.0]{by}
\acmConference[HumanSys '26]{The 4th International Workshop on Human-Centered Sensing, Modeling, and Intelligent Systems}{October 26--30, 2026}{Austin, TX, USA}
\acmBooktitle{The 4th International Workshop on Human-Centered Sensing, Modeling, and Intelligent Systems (HumanSys '26), October 26--30, 2026, Austin, TX, USA}
\acmDOI{10.1145/3842436.3843817}
\acmISBN{979-8-4007-2959-1/26/10}

\title{Will My Assistant Remember My Allergy?\\ What Personal LLM Assistants Forget When Conversation Memory Is Compressed}

\author{Lichen Zhu}
\email{lichen.zhu@duke.edu}
\affiliation{%
  \institution{Duke University}
  \city{Durham}
  \state{NC}
  \country{USA}}
\author{Yueqian Lin}
\email{y.lin@duke.edu}
\affiliation{%
  \institution{Duke University}
  \city{Durham}
  \state{NC}
  \country{USA}}
\author{Yiheng Wang}
\email{yiheng.wang@duke.edu}
\affiliation{%
  \institution{Duke University}
  \city{Durham}
  \state{NC}
  \country{USA}}
\author{Yudong Liu}
\email{yudong.liu@duke.edu}
\affiliation{%
  \institution{Duke University}
  \city{Durham}
  \state{NC}
  \country{USA}}
\author{Hai ``Helen'' Li}
\email{hai.li@duke.edu}
\affiliation{%
  \institution{Duke University}
  \city{Durham}
  \state{NC}
  \country{USA}}
\author{Yiran Chen}
\email{yiran.chen@duke.edu}
\affiliation{%
  \institution{Duke University}
  \city{Durham}
  \state{NC}
  \country{USA}}

\begin{abstract}
Personal LLM assistants (health companions, elder-care agents, accessibility
aides) are judged by what they remember about a person: a medication or an
allergy mentioned in passing and needed days later. Privacy pushes them
on-device, where a month of conversation can outgrow the model's own weights,
so an \emph{eviction policy} must decide what the cache forgets. Benchmarks
report that eviction keeps such facts at a 20\% budget, but they compress a
prompt that \emph{already contains the user's future question}, foresight no
cache-reusing assistant has. Hide the question until after compression and the
advantage vanishes: on PA-Bench, 100 assistant conversations we construct, an
allergy mentioned in passing survives to the question that needs it $0$--$1\%$
of the time, against $97\%$ with full memory. The cause is the budget, not the
scorer: none of the training-free policies we evaluate ranks the fact high
enough, and the budget that would keep it is too large to bother compressing.
A compressed cache is an inference-reuse mechanism, not a persistence layer:
safety-critical facts need an auditable episodic store alongside it, and an
interface that asks rather than invents.
\end{abstract}



\begin{teaserfigure}
\centering
\includegraphics[width=1.0\textwidth]{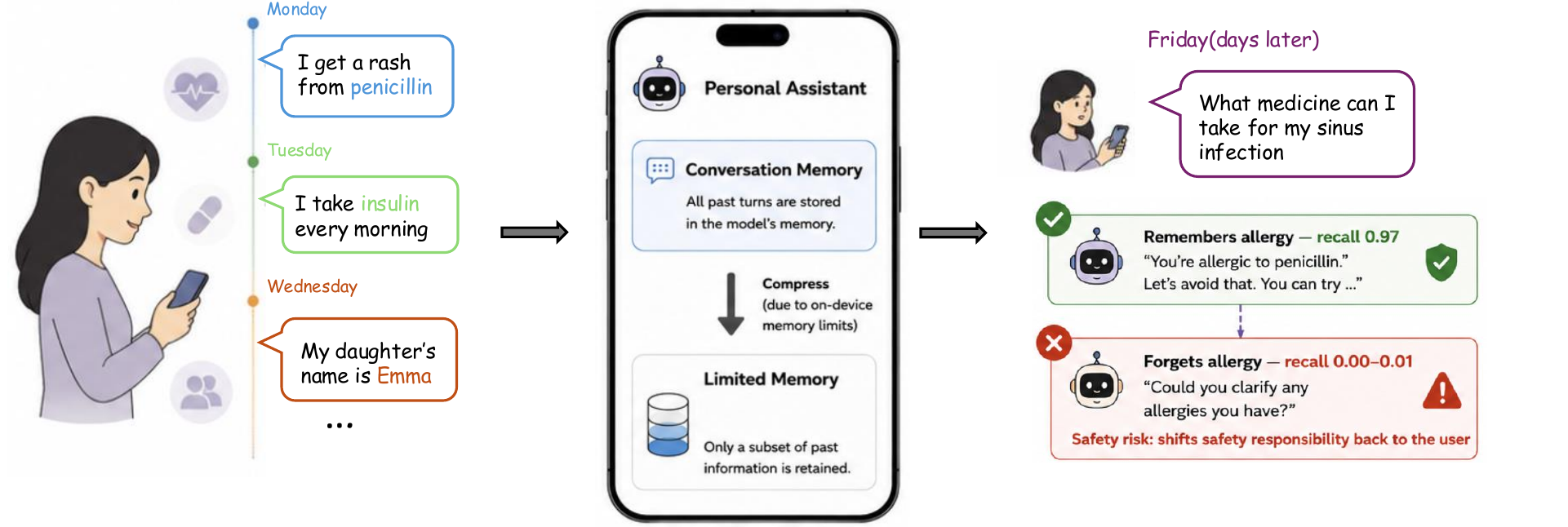}
\caption{\textbf{In deployment, memory is compressed before the question
exists.} A user discloses safety-critical facts in passing over days; the
on-device assistant compresses its memory, and only what survives is available
when the relevant question arrives. Every evaluated policy evicts
the allergy's cache blocks and forgets it. Replies are condensed from actual
model outputs, and the failure mode shown is the measured one: the compressed
assistant does not invent a drug, it asks the user to re-supply the fact
($99\%$ of failures, Section~\ref{sec:design}).}
\label{fig:pipeline}
\end{teaserfigure}

\maketitle

\section{Introduction}
\label{sec:intro}

Human-centered LLM agents are increasingly deployed to sustain a
\emph{relationship} rather than a session: health companions, elder-care
agents, and accessibility aides~\cite{laranjo2018,personalllm,melt}. Their
value accrues over time and is inherently longitudinal. On Monday, a user mentions,
in passing, that penicillin gives them a rash. On Friday, they ask what to
take for a sinus infection. Whether the assistant recalls Monday's disclosure
can critically determine whether Friday's advice is safe. When the assistant
we measure forgets, it asks rather than invents ($99\%$ of failures,
Section~\ref{sec:design}), a failure that is recoverable if the interface
surfaces it.

Two systems pressures collide over where that memory lives. \emph{Privacy
pushes the assistant on-device}~\cite{mobilellm,personalllm}. Yet on-device
memory is scarce, and the model's conversation memory, the key-value (KV)
cache holding every past turn, grows linearly with everything the user has
ever said, rivaling the model's own weights within weeks of use
(Section~\ref{sec:design}). Reconciling the two is the \emph{eviction policy}: the
mechanism deciding what the assistant \emph{forgets}~\cite{h2o,snapkv}.

A recent line of work reports, on multi-turn ``needle'' benchmarks, that
elaborate eviction scorers preserve dormant user facts at 20\% of the
cache~\cite{h2o,snapkv,defensivekv,epicache}. We set out to answer a
practitioner's question---which policy should an assistant ship?---and could
not answer it from that evidence: the benchmarks measure
\emph{query-visible, batch} compression, while a deployed assistant performs
\emph{causal, stateful} cache reuse. The gap echoes corrective
re-evaluations in recommender systems~\cite{dacrema2019}, metric
learning~\cite{musgrave2020} and pruning~\cite{blalock2020}.

The evidence fails to transfer for two reasons that compound. First, these
policies have no reference implementation for the stateful multi-turn
setting, and reimplementing them is error-prone: our own first attempts
contained two scoring errors that depress the baselines on dormant facts
(Section~\ref{sec:audit}). Second, and more fundamentally, the
standard protocol compresses a prompt that \emph{already contains the recall
question}. A cache-reusing assistant has no such foresight: on Monday, the
policy cannot see Friday's question. Hiding the question is enough to
reverse the ranking these benchmarks report (Section~\ref{sec:exp}).

Our scope is \emph{stateful} assistants that retain and incrementally
compress a bounded KV cache across turns, the regime on-device memory
pressure forces. A stateless design that re-prefills the raw
transcript \emph{after} each question may legitimately see the query, and our
critique does not apply unchanged. What it gives up is the prefill compute
that cache reuse exists to save. Our query-hidden protocol isolates the stateful
case, with one-shot selection an optimistic abstraction of incremental
eviction (Section~\ref{sec:protocol}).

\paragraph{Contributions.} This paper makes three contributions:
\begin{itemize}
\item a query-hidden evaluation protocol that compresses memory before the
user's next question exists, the constraint deployment imposes and
benchmarks omit (Sections~\ref{sec:leak}--\ref{sec:protocol}).
\item PA-Bench, 100 span-annotated assistant conversations with
safety-critical facts disclosed in passing, on which no evaluated policy
retains a single block of those facts at a 20\% budget
(Section~\ref{sec:exp}).
\item evidence that no training-free policy we evaluate ranks the dormant
fact high enough to survive an affordable budget, and the design consequence:
durable persistence is a job for an episodic store behind a repair-oriented
interface, not for cache compression (Section~\ref{sec:design}).
\end{itemize}

\section{Related Work}
\label{sec:related}

Under a bounded KV budget, eviction
selects which cached tokens survive. Policies differ in what they look at:
\emph{query-aggregate} scorers pool attention over many positions
(H2O~\cite{h2o}, DefensiveKV~\cite{defensivekv}, EpiCache~\cite{epicache}).
\emph{Window} scorers fix importance from a short recent window
(SnapKV~\cite{snapkv}). \emph{Recency} policies keep recent tokens
(StreamingLLM~\cite{streamingllm}, LRU). \emph{Future-aware} methods come
closest to our constraint: Duo\-Attention~\cite{duoattention} and
Lookahead\-KV~\cite{lookaheadkv} score
against a \emph{predicted} query, while KVzip~\cite{kvzip} compresses
\emph{query-agnos\-tically} so any later query can be reconstructed. Each needs
training, profiling, or a reconstruction pass over the full context that
stateful reuse avoids. We characterize the training-free policies shipping today
and bracket the headroom future-aware methods compete for. LoCoMo~\cite{locomo}, LongMemEval~\cite{longmemeval},
LongBench~\cite{bai2023longbench} and RULER~\cite{ruler} test long-range
recall, but most of them leak the query into compression (Section~\ref{sec:leak}).

Every personal, human-centered agent inherits the memory subsystem we
characterize: whatever it reasons over must be in its cache at answer time.
Conversational agents in healthcare are well studied~\cite{laranjo2018},
personal LLM agents push inference on-device for efficiency and
privacy~\cite{personalllm,mobilellm,melt}, and a growing line integrates them
into human-centered systems~\cite{hargpt,sensorloop}. Modern agent stacks
split memory into two roles: the KV cache accelerates inference over the
active context, while episodic or structured stores
(MemGPT~\cite{memgpt}, Mem0~\cite{mem0}, A-MEM~\cite{amem}) hold durable
facts. Our question is the boundary between the two: which
disclosed facts survive in a bounded, reused cache without foresight of the
query---what cache retention can, and cannot, be trusted to carry
(Section~\ref{sec:design}).

\section{Two Flaws in the Standard Evaluation}
\label{sec:sources}

\subsection{The benchmark leaks the future question}
\label{sec:leak}

The dominant multi-turn protocol builds \emph{one} prompt containing the whole
conversation \emph{and the recall question}, and compresses that. The eviction
policy therefore chooses what to keep while looking at the very question whose
answer it must preserve: any scorer attending to the current query (a probe
token, or SnapKV's window once it overlaps the question) trivially keeps the
queried block and scores near the full cache. A cache-reusing assistant has no
such foresight: the question that makes an early fact relevant arrives
\emph{after} that fact was compressed. Retention is being selected using
downstream information that was not available at the moment the bounded cache
had to discard.

\subsection{Reimplementation breaks the baselines}
\label{sec:audit}

Comparing policies under a query-hidden protocol first requires implementing
them faithfully for this setting. The reference implementations of SnapKV and
H2O target \emph{single-prefill} long-context compression, where the query is
legitimately included in the prompt. Neither, to our knowledge, provides a
stateful, cache-reusing variant. Any multi-turn eviction study must therefore
reimplement these policies, and that is precisely where they break. Our own
initial implementations contained two scoring errors, both systematically
depressing the baselines specifically on \emph{dormant} facts. We report them
because they are easy to make, hard to detect, and biased in a direction that
flatters whatever new method is ultimately proposed.

In our SnapKV implementation, a past token was ranked by how similar its
\emph{key} was to recent keys, rather than by how much attention the recent
tokens actually \emph{pay} it, which is what SnapKV~\cite{snapkv} specifies. The
substitute proxy ignores rotary position encoding and degrades across long
position gaps, exactly where a dormant fact lives. Correcting it raises
SnapKV's mean recall, pooled over both planted facts in this audit harness,
from $0.12$ to $0.94$. With the faulty scorer, sweeping the observation
window from $32$ to $1024$ tokens left recall at $0.000$ throughout: the
fault was the scorer, not the window size.

Our H2O implementation made a different mistake. H2O~\cite{h2o} keeps tokens
whose attention, accumulated over \emph{all} query positions, is highest. Our
code approximated that with a \emph{single} appended probe token, collapsing a
lifetime of accumulated attention into one snapshot. Restoring the accumulation
changes H2O qualitatively: it becomes the only policy that recalls anything
under a query-hidden protocol.

The published SnapKV and H2O results are not in question: in their
single-prefill setting, with the query present, the specification and the
reported behaviour agree, and our corrected code reproduces them (SnapKV
$0.975$, query-visible). The narrower point is the one that matters to an
assistant builder: moving these policies into a cache-reusing assistant is
easy to get wrong, and the mistakes land on exactly the dormant facts it needs
to keep. A shared, query-hidden harness is the remedy, and
Section~\ref{sec:protocol} specifies ours.

\section{A Leak-Free Evaluation Protocol}
\label{sec:protocol}

Dormancy is what makes this setting hard: a query-blind scorer has one
signal, the attention a fact drew \emph{when it was said}. Ranking by attention from the
\emph{current} position (a probe, or SnapKV's window once past the fact)
scores a dormant fact near zero by construction, so accumulating attention
over a fact's lifetime (H2O) is the only query-blind signal we test with a chance
(Section~\ref{sec:exp_budget}).

We reimplement SnapKV with
query-key attention over its observation window, H2O with cumulative
attention over all query positions, plus StreamingLLM, LRU, a single-position
probe, and a two-tier INT4 demotion variant. All run at a common block
granularity under \emph{matched budgets}: each keeps exactly $B$ blocks, with
the attention sinks and the recency window counted \emph{inside} $B$, so no
policy gets extra capacity, an accounting that matters far more than it looks
(Section~\ref{sec:exp_budget}).

The two protocols we compare differ in one line of the harness: what the
scorer may see. \emph{Query-visible} (the standard protocol) compresses a
prefill that already includes the recall question. \emph{Query-hidden} (ours)
compresses the conversation \emph{before} the question exists, then reuses that
cache to answer: the question is prefilled onto the surviving blocks at their
original positions, with no re-scoring. Formally, with tokens $x_{1:T}$,
question at $[q_s,q_e]$ and answer after $a_s$, the former scores on
$\mathrm{KV}(x_{1:a_s-1})$, the latter on $\mathrm{KV}(x_{1:q_s-1})$, decoding
from the retained blocks plus $x_{q_s:a_s-1}$. Both protocols share every
other line of the harness---model, budgets, block granularity, decoding---so
any recall difference is attributable to what the scorer saw. Compression is one-shot at
$q_s$, an \emph{optimistic} abstraction of incremental per-turn eviction: a
policy that cannot keep a fact under one-shot selection is unlikely to keep it
under repeated eviction either.

\section{Results and Analysis}
\label{sec:exp}

We use a controlled multi-turn needle benchmark: 30-round
conversations ($\sim$2--3K tokens) with an early and a mid fact recalled at the
end. Runs use $n{=}100$, Qwen2.5-7B-Instruct, a 20\% budget, block size 8, bf16, and Wilson
95\% CIs. Full-memory recall is $1.0$ on larger models, so the bottleneck is
eviction, not capability. We also evaluate on LongBench, RULER, LongMemEval and
LoCoMo, and replicate on Llama-3.2-3B and Qwen3-8B.

\subsection{Hide the question and the best scorer keeps nothing}
\label{sec:exp_main}

With the question visible, the scorers that attend to it dominate (SnapKV
$0.975$, probe $0.945$, $\approx$ full $0.93$), simply by keeping the queried
block. Hide the question and that ranking collapses (Table~\ref{tab:main},
corrected and budget-matched): both query-specific scorers fall to $0.00$, and
only cumulative H2O still recalls anything, and only on the synthetic needle
($0.17$, mostly the early fact); on PA-Bench even H2O falls to $0.00$ and keeps
none of the fact's blocks (Ret.). The full cache still scores $0.92$, so the
model can answer. The eviction policy is what fails. The collapse is not specific to one model: across three families
(Table~\ref{tab:xmodel}), query-specific scorers that dominate query-visible
fall to near-zero once the question is hidden. Qwen3-8B is a telling case: its
prefill attention is diffuse enough that query-specific scorers are weak
\emph{even query-visible}, yet the query-hidden ordering is the same, and H2O
remains the sole survivor throughout.

\begin{table}[t]
\centering
\caption{Recall on the synthetic needle and on PA-Bench, plus
query-hidden retention of the fact's KV blocks (Ret.). Q-Vis.\ is the
query-visible protocol, Q-Hid.\ the query-hidden one (Section~\ref{sec:protocol}).
Qwen2.5-7B, matched 20\% budget, $n{=}100$ instances. Wilson 95\% CI
half-widths $\le 0.07$ on the $2n{=}200$ pooled fact outcomes per cell.
Full-cache differences between protocols ($0.93$/$0.92$, $0.96$/$0.97$) are
within sampling noise.}
\label{tab:main}
\small
\setlength{\tabcolsep}{3.2pt}
\begin{tabular}{lcc|ccc}
\toprule
 & \multicolumn{2}{c|}{Synthetic needle} & \multicolumn{3}{c}{PA-Bench (assistant-style)} \\
Policy & Q-Vis. & Q-Hid. & Q-Vis. & Q-Hid. & Ret. \\
\midrule
Full cache   & 0.93 & 0.92 & 0.96 & 0.97 & 1.00 \\
SnapKV       & \textbf{0.975} & 0.00 & 0.83 & 0.01 & 0.00 \\
Probe        & 0.945 & 0.00 & \textbf{0.87} & 0.00 & 0.00 \\
H2O          & 0.27 & \textbf{0.17} & 0.085 & 0.00 & 0.00 \\
StreamingLLM & 0.08 & 0.00 & 0.055 & 0.00 & 0.00 \\
LRU          & 0.035 & 0.00 & 0.09 & 0.00 & 0.00 \\
\midrule
\rowcolor{keep_green!12}
Query-oracle & --- & 0.975 & --- & --- & --- \\
\bottomrule
\end{tabular}
\end{table}

\begin{table}[t]
\centering
\caption{The collapse is cross-family. Needle recall, 20\% budget, reported as
query-visible\,$\to$\,query-hidden. Query-specific scorers (SnapKV, probe)
collapse to near-zero once the question is hidden. Only aggregate H2O retains a
residual, on every model.}
\label{tab:xmodel}
\small
\setlength{\tabcolsep}{5pt}
\begin{tabular}{lccc}
\toprule
Model & SnapKV & Probe & H2O \\
\midrule
Qwen2.5-7B   & $0.975\!\to\!0.00$ & $0.945\!\to\!0.00$ & $0.27\!\to\!0.17$ \\
Llama-3.2-3B & $0.99\!\to\!0.035$ & $0.915\!\to\!0.00$ & $0.285\!\to\!0.175$ \\
Qwen3-8B     & $0.07\!\to\!0.05$ & $0.275\!\to\!0.00$ & $0.10\!\to\!0.085$ \\
\bottomrule
\end{tabular}
\end{table}

Eviction itself is not broken. Dormant recall is. On four established
long-context benchmarks the corrected policies are statistically
indistinguishable from one another \emph{and from the full cache}: they lose
nothing when the evidence is still salient as the question arrives: on
LongBench QA ($n{=}200$), SnapKV F1 is $0.156$ against the probe's $0.151$
(bootstrap $p{=}0.63$); on RULER, H2O scores $0.89$ against the probe's
$0.88$; on LongMemEval, $0.129$ against $0.121$; and LoCoMo is saturated at
its 8K cap (full cache F1 $=0.11$), so we read no ordering from it. Our
finding is specific to the
case an assistant lives in: a fact that went \emph{dormant} before the question
existed. Nothing here argues against attention-guided eviction. With the query
in hand, a corrected SnapKV is an excellent scorer ($0.975$). The problem is
that a cache-reusing assistant never has it.

\subsection{No evaluated policy keeps the allergy in assistant conversations}
\label{sec:exp_pa}

To test this in assistant-style traffic, we built \emph{PA-Bench}, a
\emph{span-annotated} benchmark designed for controlled diagnosis rather than
ecological realism: $n{=}100$ thirty-round
conversations in which the user issues everyday requests (reminders, weather,
shopping lists) and states two safety-critical facts in passing (e.g., a drug
allergy, a medication, an emergency contact) from templated pools of
clinically-plausible values, with the recall query at the end. Each
conversation plants fact A at round 3 and fact B at round 16 of the 30
rounds, with values drawn from pools of distinctive strings so that
case-insensitive substring scoring is unambiguous; Table~\ref{tab:patypes}
lists the six fact types with their disclosures and recall queries. All
values are synthetic, so no real personal health information is collected or
processed in this study. Since the
spans are planted, the exact KV blocks holding each fact are known, and
beyond recall we log \emph{fact retention}: the fraction of those blocks a
policy keeps.

\begin{table}[t]
\centering
\caption{The six PA-Bench fact types. Each conversation plants two of them,
disclosed in passing amid everyday requests, and the recall query arrives
only at the end, after compression.}
\label{tab:patypes}
\small
\renewcommand{\arraystretch}{1.15}
\setlength{\tabcolsep}{3pt}
\begin{tabular}{@{}p{0.19\columnwidth}p{0.43\columnwidth}p{0.30\columnwidth}@{}}
\toprule
Type & Disclosure (in passing) & Recall query \\
\midrule
Allergy & ``by the way, I'm allergic to \{drug\}---I get a bad rash from it.'' & Which medication am I allergic to? \\
Medication & ``my doctor put me on \{med\} for my blood pressure, one tablet every morning.'' & What medication do I take every morning? \\
Doctor & ``my new cardiologist is Dr.~\{name\}, at the clinic on 5th Street.'' & What is my cardiologist's name? \\
Emergency & ``if anything ever happens to me, my emergency contact is my daughter \{name\}.'' & Who is my emergency contact? \\
Diet & ``the dietitian said I must avoid \{food\} completely because of my condition.'' & Which food must I avoid completely? \\
Appointment & ``my follow-up appointment with the specialist is on \{date\}, please remember that.'' & When is my follow-up appointment with the specialist? \\
\bottomrule
\end{tabular}
\end{table}

Table~\ref{tab:main} (right): query-hidden, the collapse is \emph{total}. Even
cumulative H2O, the only survivor on the synthetic needle ($0.17$), drops to
$0.00$, and no compressed policy keeps a single block of the sensitive fact
(retention $0.00$, versus $0.86$ when the query is leaked). The mechanism shows
in the query-visible column: amid uniformly salient requests, a passing health
disclosure attracts no distinctive aggregate attention (H2O $0.27$ synthetic
$\to$ $0.085$ here). Synthetic needles, which plant facts with emphatic framing
amid bland filler, \emph{overstate} what query-blind compression preserves. The
zeros are not decode failures: generations stay coherent (the assistant asks
the user to re-supply the fact), and H2O ranks the fact's blocks 16--59 of 161
(probe: 96--128), above median, yet outside the $\sim$7 score-driven slots left
once sinks and recency take the rest of the 20\% budget.

\subsection{The allergy returns only when compression stops paying}
\label{sec:exp_budget}

\begin{figure*}[t]
\centering
\includegraphics[width=0.98\textwidth]{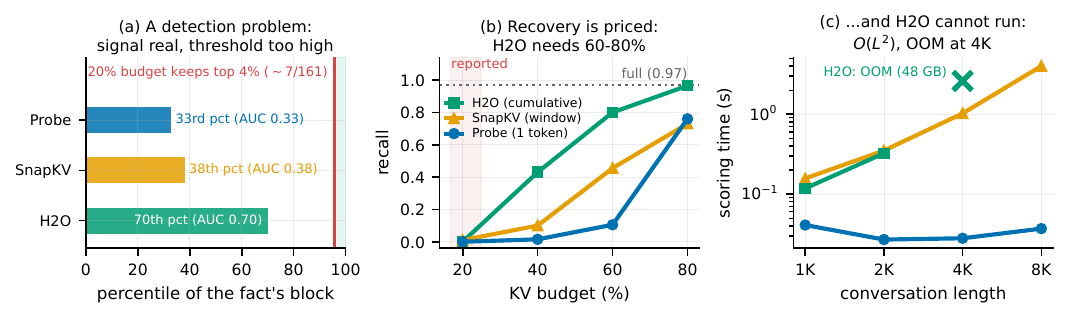}
\caption{\textbf{Dormant recall is not undetectable, just unaffordable}
(PA-Bench, query-hidden).
\textbf{(a)} H2O ranks the allergy block above most others (AUC $0.70$), but a
20\% budget keeps too few to reach it.
\textbf{(b)} Recall returns only as the budget grows ($0.43$ at 40\%, $0.80$
at 60\%), tracking block retention.
\textbf{(c)} H2O retains the most but is the costliest to score: a naive
implementation OOMs past 4K tokens, though an efficient kernel would accumulate
column-sums instead.}
\label{fig:results}
\end{figure*}

How much cache does dormant retention actually need? We swept the budget from
20\% to 80\% (Fig.~\ref{fig:results}). Read as detection, the
mechanism is precise: H2O ranks the health fact at the 70th percentile of all
blocks (AUC $0.70$), a real if modest signal, while the window and probe
scorers fare worse (38th, 33rd), since a scorer asking ``what matters
\emph{now}'' cannot see a fact that went dormant turns ago. But the decisive
quantity is the threshold, not detectability. A 20\% budget on our
$\sim$161-block conversations keeps 32 blocks, of which sinks and recency
reserve $\sim$25, so the scorer chooses only $\sim$7. A nominal 20\%
cache is a 4\% \emph{scoring} cache admitting only blocks above the 96th
percentile. The scorer does rank the fact above median. The budget simply never
reaches it (Table~\ref{tab:budget}). Its blocks sit at ranks 16--59 in H2O's
scores (Section~\ref{sec:exp_pa}), so the $\sim$7 slots at 20\% miss it
entirely, 39 at 40\% reach into the span, 71 at 60\% cover it, and 103 at
80\% clear it. The fact
comes back when the budget grows past its rank, not when the scorer improves.

\begin{table}[t]
\centering
\caption{The budget sweep, read as capacity meeting rank (PA-Bench,
query-hidden, H2O, $\sim$161-block conversations). Sinks and recency reserve
$\sim$25 blocks at every budget, and the fact's blocks sit at ranks 16--59.
Recall returns as the score-driven slots reach that span.}
\label{tab:budget}
\small
\begin{tabular}{cccc}
\toprule
Budget & Kept blocks & Score-driven slots & Recall \\
\midrule
20\% & 32 & $\sim$7  & 0.00 \\
40\% & 64 & 39 & 0.43 \\
60\% & 96 & 71 & 0.80 \\
80\% & 128 & 103 & 0.965 \\
\bottomrule
\end{tabular}
\end{table}

A bigger budget does not resolve the tension. The ranking holds at every budget
(H2O $>$ SnapKV $>$ probe), as the query-blind argument predicts, but the fact
only comes back once the cache is barely compressed: the $2.4\times$ decode
speedup comes from the same memory that has to be \emph{dropped}, and the one
policy that recovers the fact, H2O, is also the most expensive to score
(Fig.~\ref{fig:results}c): scoring needs the attention every key received, which
a naive implementation forms as a full $O(L^2)$ matrix, exhausting a 48\,GB GPU
past 4K tokens. That is an implementation limit, not a fundamental one: an
efficient H2O accumulates the same column-sums without ever forming the
matrix. But it makes the point that the cheapest-to-score policies (probe,
window) are exactly the ones that retain nothing. Recovering a dormant fact without the query is
possible, then, but only by keeping enough cache that compression no longer
pays.

At an affordable budget, only knowing the question recovers the fact. No
cheaper signal we test substitutes. A SnapKV scorer allowed to see the question
reaches $0.975$, an \emph{oracle reference} for what future-aware methods
(DuoAttention~\cite{duoattention}, LookaheadKV~\cite{lookaheadkv},
KVzip~\cite{kvzip}) compete for. We omit them as they need training, profiling,
or a full-context reconstruction pass. Generic prediction does not suffice: one LLM-generated candidate
question as the SnapKV window recovers only $0.09$, since generic guesses track
recency, not the dormant fact. Nor does spending the saved bytes on
precision: a byte-fair INT4 demotion tier moves recall from $0.00$ only to
$0.01$. The information the scorer lacks is not in the cache; it is in the
future.

\section{Implications for Assistant Builders}
\label{sec:design}

A compressed cache is an inference-reuse mechanism, not the assistant's
memory system. Conversation
memory is a long-lived assistant's dominant, fastest-growing cost
(Table~\ref{tab:sys}): at $56$\,KB/token it passes the model's own INT4
weights ($3.8$\,GB) within a month of use, so compression is mandatory, and it
pays (up to $2.4\times$ faster decoding). But a compressed cache fails to hold
what matters (retention $0.00$, query-hidden), so safety-critical facts need
an explicit episodic store~\cite{memgpt} alongside it, re-injected on demand.
We measure one directly, as a proof of concept: a single untuned prompt asking
the same model to extract durable facts, re-injected at query time, recovers
the allergy in $0.51$ of cases, against $0.00$ for any compressed cache. The
store relocates the problem more than it solves it: its recall equals its
extraction coverage ($0.51$), and the compressed cache cannot
back-stop the misses, so the bottleneck moves from retaining the fact to
recognizing it as worth storing, a more tractable and more human-centered
target: a small structured store is one the user can inspect and correct,
and, unlike a cache, one whose contents can be deliberately deleted.

\begin{table}[t]
\centering
\caption{The systems cost of conversation memory (Qwen2.5-7B, bf16, measured
on an L40S, with per-token arithmetic that is hardware-independent). Memory
grows past the model's own INT4 weights ($3.8$\,GB) within a month of use, and
compacting the cache speeds decoding because decoding is memory-bound.}
\label{tab:sys}
\small
\begin{tabular}{lcc}
\toprule
Context & KV cache & Decode speedup (20\% cache) \\
\midrule
8K                  & $0.47$\,GB & $1.3\times$ \\
16K                 & $0.94$\,GB & $1.8\times$ \\
32K (one session)   & $1.88$\,GB & $2.4\times$ \\
100K ($\sim$a month) & $5.7$\,GB  & --- \\
\bottomrule
\end{tabular}
\end{table}

Report the scoring budget, not the nominal one. Sinks and recency consume most
of it before scoring begins, leaving $\sim$7 of the 32 kept blocks at 20\%, a
4\% \emph{scoring} cache (Section~\ref{sec:exp_budget}); a policy chosen on
the nominal budget silently under-retains. The same retention curve is also a
disclosure-persistence curve: at 20\% query-hidden \emph{no} evaluated policy
kept the health fact, but under the leaky protocol query-aware scorers kept
$86\%$, so ``the cache probably evicted it'' is not a deletion guarantee.

Design the interface for repair, because forgetting surfaces as
\emph{omission}, not \emph{confabulation}: on PA-Bench $99\%$ ($196/198$) of
failures were repair requests---the assistant asked the user to re-supply the
fact---and \emph{not one} confabulated a different drug or contact. That benign mode is
the model's honesty, not a policy guarantee. Surfacing the gap (``I have no record of medication
allergies, is that right?'') turns a silent safety failure into a recoverable
turn, showing memory uncertainty to the user instead of hiding it in the
cache. The same interface can make the episodic store legible at write time:
an assistant that confirms a disclosure as it stores it (``noted---allergic
to penicillin'') gives the user a chance to correct the record on the spot.

Personalization is the remaining lever. No training-free policy we evaluate
is both effective without foresight and deployable, but an assistant knows its
user:
routines and a calendar make the next question partially predictable (a
refill reminder scheduled for Friday makes a medication question likely), and
near-full recall is available once it is known (Section~\ref{sec:exp_budget}),
so an on-device predictor tuned to one user's history is the most promising way
to close the gap without buying memory. For evaluators: never let the recall
query enter the compression decision, and report the full-memory reference at
every budget.

\section{Future Work}
\label{sec:future}

Our study leaves clear next steps. PA-Bench is templated, so the collapse
should next be checked on span-annotated real conversations, where
disclosures are messier, paraphrased, and revisited across turns, and
evaluation should widen to complete on-device memory pipelines, whose
retrieval accuracy, latency, and energy trade against compressed-cache reuse.
Our single-mask, block-granular budgeting scopes the results to training-free
policies and likely understates per-head methods; finer granularity may
recover part of the gap, though it does not change what a query-blind scorer
can see. One-shot selection is an optimistic abstraction, so measuring how
repeated per-turn eviction compounds the loss is a natural extension. The
protocol also extends beyond text: image and sensor context carries far
larger KV footprints, and whether its eviction obeys the same query-blind
limits is untested here. Future-aware policies
(DuoAttention~\cite{duoattention}, LookaheadKV~\cite{lookaheadkv},
KVzip~\cite{kvzip}) may close the gap, at the cost of training, profiling, or a
reconstruction pass. Whether a
personalized on-device predictor (Section~\ref{sec:design}) can close the gap
without that cost is the question we find most open.

\section{Conclusions}
\label{sec:conclusion}

At deployable budgets, all training-free policies we evaluate forget dormant
safety-critical facts: on PA-Bench, hiding the question turns query-visible
recall of $0.83$--$0.87$ into $0.00$--$0.01$, against $0.97$ with full
memory. \emph{Dormant-fact recall is not impossible—it is
impractical under the compression budgets we study.} A compressed cache should
be treated as an inference-reuse mechanism; safety-critical facts should live
in an auditable episodic memory alongside it, not be entrusted to eviction.
Whether the assistant remembers the allergy should be a measured property of
the system, not an assumption inherited from a leaky benchmark.

\begin{acks}
This work was supported in part by the National Science Foundation (NSF)
under Grant No.~2112562 and the Army Research Office (ARO) under Grant
No.~W911NF-23-2-0224.
\end{acks}

\bibliographystyle{ACM-Reference-Format}
\bibliography{references}

\end{document}